# AWGN channel analysis of terminated LDPC convolutional codes

David G. M. Mitchell*, Michael Lentmaier†, and Daniel J. Costello, Jr.*
*Dept. of Electrical Engineering, University of Notre Dame, Notre Dame, Indiana, USA,
{dcostel1, david.mitchell}@nd.edu
†Vodafone Chair Mobile Communications Systems, Dresden University of Technology, Dresden, Germany,
michael.lentmaier@ifn.et.tu-dresden.de

***Abstract*—It has previously been shown that ensembles of terminated protograph-based low-density parity-check (LDPC) convolutional codes have a typical minimum distance that grows linearly with block length and that they are capable of achieving capacity approaching iterative decoding thresholds on the binary erasure channel (BEC). In this paper, we review a recent result that the dramatic threshold improvement obtained by terminating LDPC convolutional codes extends to the additive white Gaussian noise (AWGN) channel. Also, using a (3,6)-regular protograph-based LDPC convolutional code ensemble as an example, we perform an asymptotic trapping set analysis of terminated LDPC convolutional code ensembles. In addition to capacity approaching iterative decoding thresholds and linearly growing minimum distance, we find that the smallest non-empty trapping set of a terminated ensemble grows linearly with block length.**

## I. Introduction

Ensembles of low-density parity-check (LDPC) block codes can be obtained by terminating LDPC convolutional code ensembles [1], [2]. The slight irregularity resulting from the termination of the convolutional codes has been shown to lead to substantially better belief propagation (BP) decoding thresholds compared to corresponding block code ensembles. More recently, it has been proven analytically for the binary erasure channel (BEC) that the BP decoding thresholds of some slightly modified regular LDPC convolutional code ensembles approach the maximum a posteriori probability (MAP) decoding thresholds of the corresponding LDPC block code ensembles [3]. Figure 1 displays the simulated performance of terminated $(3,6)$-regular and $(4,8)$-regular LDPC convolutional codes over the additive white Gaussian noise (AWGN) channel and serves as a demonstration of the capabilities of these codes and as motivation for the results presented in this paper. Here, the termination length has been chosen such that the code rate is $R = 0.49$. We note in particular that as $J$ increases, the thresholds and corresponding waterfall performance of the simulated codes improves.

In addition to this excellent threshold performance, it can also be shown that the minimum distance typical of most members of these terminated LDPC convolutional code ensembles grows linearly with the block length as the block length tends to infinity, i.e., they are *asymptotically good* [4]. A large minimum distance growth rate indicates that codes drawn from the ensemble should have a low error floor under maximum likelihood (ML) decoding. However, when sub-optimal decoding methods are employed, there are other factors that affect the performance of a code. For example, it has been shown that so-called 'trapping sets' are a significant factor affecting decoding failures of LDPC codes over the AWGN channel with iterative message-passing decoding. Trapping sets, graphical sub-structures existing in the Tanner graph of LDPC codes, were first studied in [5]. Known initially as *near-codewords*, they were used to analyse the performance of LDPC codes in the error floor, or high signal-to-noise ratio (SNR) region, of the bit error rate (BER) curve. In [6], Richardson developed these concepts and proposed a two-stage technique to predict the error floor performance of LDPC codes based on trapping sets, and asymptotic results on average trapping set distributions for both regular and irregular LDPC block code ensembles appeared in [7].

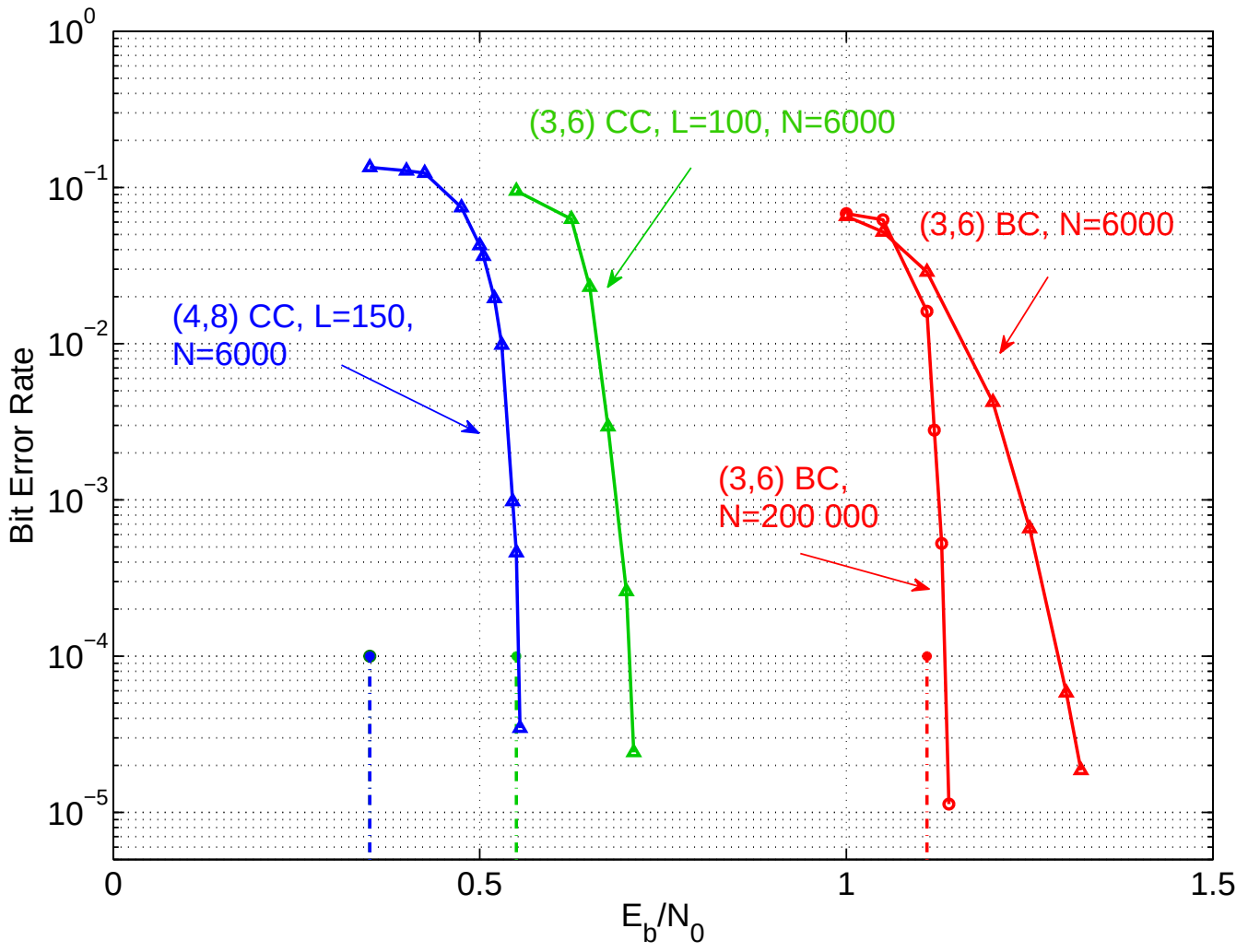

Fig. 1: AWGN channel performance of terminated (3,6)-regular and (4,8)-regular LDPC convolutional codes with lifting factor $N = 6000$ and rate $R = 0.49$. For comparion, (3,6)-regular LDPC block codes with $N = 6000$ and $N = 200000$ are also shown.

In this paper, we perform an AWGN channel analysis of terminated LDPC convolutional codes. We begin in Section III by briefly reviewing a recent result that the dramatic threshold improvement obtained by terminating LDPC convolutional codes on the BEC also extends to the AWGN channel. This result is demonstrated for a variety of asymptotically good, rate $R = 1/2$ $(J, 2J)$-regular LDPC convolutional code ensembles and for a rate $R = 1/2$ irregular LDPC convolutional code ensemble based on the accumulate-repeat-jagged-accumulate (ARJA) block protograph [8]. In Section IV, using a (3,6)-regular protograph-based LDPC convolutional code ensemble as an example, we perform an asymptotic trapping set analysis of terminated LDPC convolutional code ensembles. Here, using techniques developed by Abu-Surra, Ryan, and Divsalar [9], asymptotic methods are used to calculate a lower bound on the trapping set numbers of terminated $(3, 6)$-regular LDPC convolutional code ensembles. These trapping set numbers define the size of the smallest, non-empty trapping sets in an ensemble. Concluding remarks are given in Section V.

## II. Constructing protograph-based LDPC convolutional codes

A protograph [10] is a small bipartite graph $B = (V, C, E)$ that connects a set of $n_v$ variable nodes $V = \{v_0, \dots, v_{n_v-1}\}$ to a set of $n_c$ check nodes $C = \{c_0, \dots, c_{n_c-1}\}$ by a set of edges $E$. The protograph can be represented by a parity-check or *base* biadjacency matrix $\mathbf{B}$, where $B_{x,y}$ is taken to be the number of edges connecting variable node $v_y$ to check node $c_x$. The parity-check matrix $\mathbf{H}$ of a protograph-based LDPC block code can be created by replacing each non-zero entry in $\mathbf{B}$ by a sum of $B_{x,y}$ permutation matrices of size $N$ and a zero entry by the $N \times N$ all-zero matrix. In graphical terms, this can be viewed as taking an $N$-fold graph cover [11] or "lifting" of the protograph. It is an important feature of this construction that each lifted code inherits the degree distribution and graph neigbourhood structure of the protograph. The ensemble of protograph-based LDPC codes with block length $n = Nn_v$ is defined by the set of matrices $\mathbf{H}$ that can be derived from a given protograph by all possible combinations of $N \times N$ permutation matrices.

### A. Convolutional protographs

An ensemble of unterminated LDPC convolutional codes can be described by means of a *convolutional protograph* [1] with base matrix

$$\mathbf{B}_{[-\infty,\infty]} = \begin{bmatrix} \ddots & & \ddots & & & \\ & \mathbf{B}_{m_s} & \cdots & \mathbf{B}_0 & & \\ & & \ddots & & \ddots & \\ & & & \mathbf{B}_{m_s} & \cdots & \mathbf{B}_0 \\ & & & & \ddots & & \ddots \end{bmatrix}, \quad (1)$$

where $m_s$ denotes the syndrome former memory of the convolutional codes and the $b_c \times b_v$ *component base matrices* $\mathbf{B}_i$, $i = 0, \dots, m_s$, represent the edge connections from the $b_v$ variable nodes at time $t$ to the $b_c$ check nodes at time $t + i$. Starting from the base matrix $\mathbf{B}$ of a block code ensemble, one can construct LDPC convolutional code ensembles that maintain the same degree distribution and structure as the original ensemble. This is achieved by an *edge spreading* procedure (see [1] for details) that divides the edges from each variable node in the base matrix $\mathbf{B}$ among $m_s + 1$ component base matrices $\mathbf{B}_i$, $i = 0, \dots, m_s$, such that the condition $\mathbf{B}_0 + \mathbf{B}_1 + \cdots + \mathbf{B}_{m_s} = \mathbf{B}$ is satisfied. This ensures that the computation trees of the convolutional code ensemble are equal to those of the original block code ensemble defined by $\mathbf{B}$. An ensemble of (in general) time-varying LDPC convolutional codes can then be formed from $\mathbf{B}_{[-\infty,\infty]}$ using the protograph construction method based on $N \times N$ permutation matrices described above.

For example, a (3,6)-regular LDPC convolutional ensemble with $m_s = 2$ can be formed from the block base matrix $\mathbf{B} = [\,3\;\;3\,]$ by defining the component base matrices

$$\mathbf{B}_0 = \begin{bmatrix} 1 & 1 \end{bmatrix} = \mathbf{B}_1 = \mathbf{B}_2\,, \quad (2)$$

with corresponding convolutional base matrix

$$\mathbf{B}_{[-\infty,\infty]} = \begin{bmatrix} \ddots & & \ddots & & \ddots & & & & & \\ 1 & 1 & 1 & 1 & 1 & 1 & & & & \\ & & 1 & 1 & 1 & 1 & 1 & 1 & & \\ & & & & 1 & 1 & 1 & 1 & 1 & 1 \\ & & & & & \ddots & & \ddots & & \ddots \end{bmatrix}.$$

### B. Terminated LDPC convolutional code ensembles

Suppose that we start the convolutional code with parity-check matrix defined in $(1)$ at time $t = 0$ and terminate it after $L$ time instants. The resulting finite-length base matrix is then given by

$$\mathbf{B}_{[0,L-1]} = \begin{bmatrix} \mathbf{B}_0 & & \\ \vdots & \ddots & \\ \mathbf{B}_{m_s} & & \mathbf{B}_0 \\ & \ddots & \vdots \\ & & \mathbf{B}_{m_s} \end{bmatrix}_{(L+m_s)b_c \times Lb_v}. \quad (3)$$

The matrix $\mathbf{B}_{[0,L-1]}$ can be considered as the base matrix of a terminated protograph-based LDPC convolutional code ensemble. Termination in this fashion results in a rate loss. Without puncturing, the design rate $R_L$ of the terminated code ensemble is equal to

$$R_L = 1 - \left(\frac{L + m_s}{L}\right)\frac{b_c}{b_v} = 1 - \left(\frac{L + m_s}{L}\right)(1 - R)\,,$$

where $R = 1 - Nb_c/Nb_v = 1 - b_c/b_v$ is the rate of the unterminated convolutional code ensemble. Note that, as the termination factor $L$ increases, the rate increases and approaches the rate of the unterminated convolutional code ensemble.

## III. Iterative decoding thresholds and minimum distance growth rates of terminated ensembles

Terminated LDPC convolutional code ensembles have been observed to display BP thresholds that approach the MAP decoding threshold of the corresponding block code ensembles for the BEC as the termination factor $L$ tends to infinity [1], [12], and recently this has been proven analytically for $(J, K)$-regular ensembles [3]. In this section, we review a more recent result that the dramatic threshold improvement obtained by terminating LDPC convolutional codes also extends to the AWGN channel [13], [14]. Since exact density evolution is far more complex for the AWGN channel than for the BEC, we make use of the reciprocal channel approximation (RCA) technique introduced in [15], which has been successfully applied to the analysis of protograph ensembles in [8]. With this approach, the calculation of approximate AWGN channel thresholds for large protographs becomes feasible with reasonable accuracy. Typical minimum distance growth rates for protograph-based terminated LDPC convolutional code ensembles can be calculated using the techniques presented by Divsalar [16].

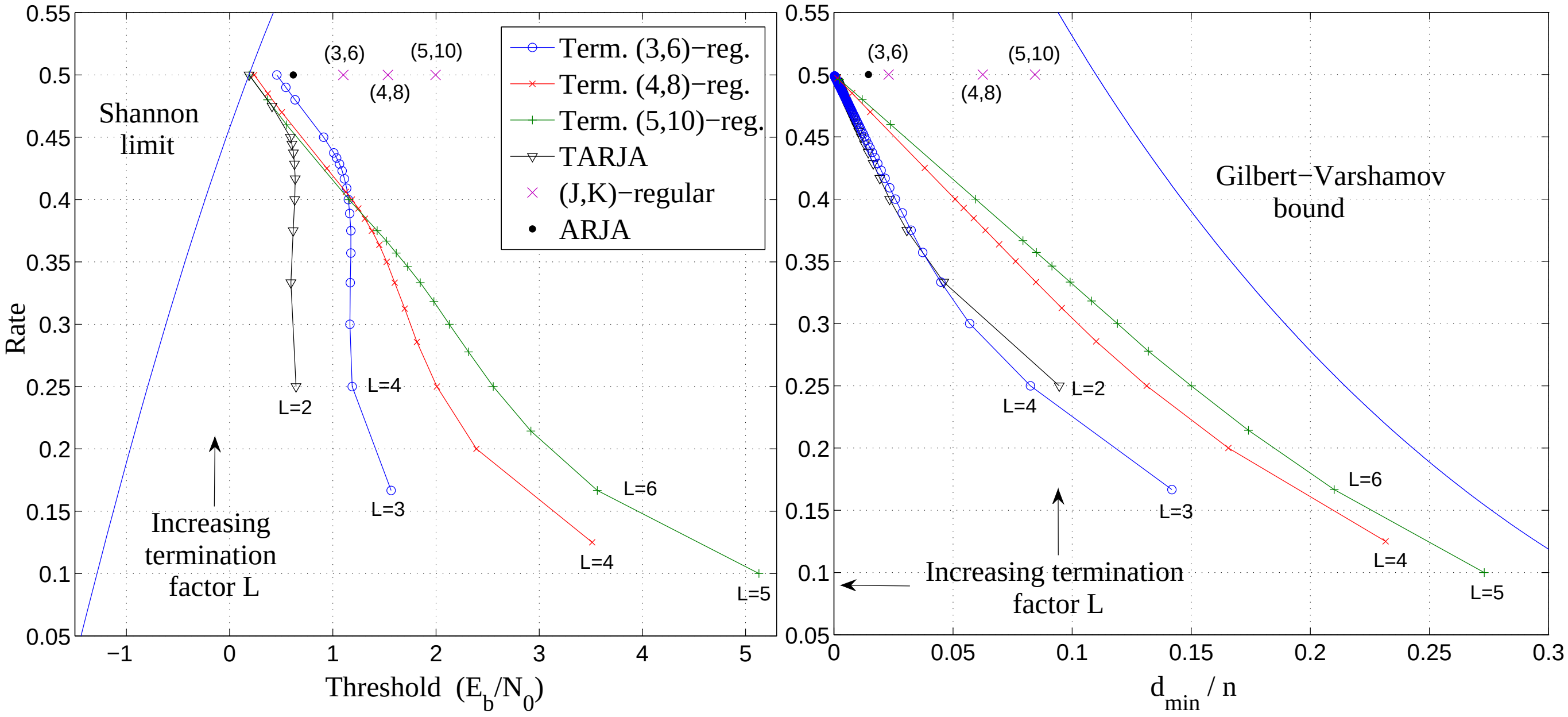

Fig. 2: AWGN thresholds and typical minimum distance growth rates for families of terminated $(J, 2J)$-regular LDPC convolutional code ensembles, terminated ARJA LDPC convolutional code ensembles, and associated LDPC block code ensembles.

Figure 2 displays the calculated AWGN thresholds and minimum distance growth rates for the terminated LDPC convolutional code ensembles and the corresponding LDPC block code ensembles. The $(J, 2J)$-regular terminated LDPC convolutional base matrices are given by (3), where the $b_c \times b_v = 1 \times 2$ component submatrices $\mathbf{B}_i = [1\ 1]$, $i = 0, \ldots, m_s$, and $m_s = J - 1$. The terminated convolutional ARJA (TARJA) component submatrices $\mathbf{B}_0$ and $\mathbf{B}_1$ of size $b_c \times b_v = 3 \times 5$ are given as follows:

$$\mathbf{B}_0 = \begin{bmatrix} 1 & 2 & 0 & 0 & 0 \\ 0 & 1 & 1 & 1 & 0 \\ 0 & 0 & 1 & 0 & 2 \end{bmatrix} \text{ and } \mathbf{B}_1 = \begin{bmatrix} 0 & 0 & 0 & 0 & 0 \\ 0 & 2 & 0 & 0 & 1 \\ 0 & 1 & 1 & 1 & 0 \end{bmatrix},$$

where we note that

$$\mathbf{B}_0 + \mathbf{B}_1 = \begin{bmatrix} 1 & 2 & 0 & 0 & 0 \\ 0 & 3 & 1 & 1 & 1 \\ 0 & 1 & 2 & 1 & 2 \end{bmatrix} = \mathbf{B},$$

the block base matrix of the ARJA ensemble, and that the variable node associated with column 2 should be punctured.

We see for the terminated convolutional ensembles that, as $L$ increases, the ensemble design rate increases (approaching $R = 1/2$ asymptotically) and the thresholds approach the Shannon limit. However, for increasing $L$, the asymptotic minimum distance growth rates $\delta_{min}^{(L)}$ of the ensembles decrease. This presents the code designer with a trade-off between distance growth rate and threshold, along with a variety of achieveable code rates $R_L$. We observe that the $(J, 2J)$-regular block code ensemble thresholds worsen as we increase $J$. Figure 2 shows that this is also the case for the terminated $(J, 2J)$-regular convolutional code families for small termination factors $L$. Thus, by increasing $J$ (and hence decoding complexity), we obtain a more pronounced trade-off between distance growth rates and threshold for small values of $L$. However, as the termination factor $L$ increases, we observe that the threshold of the terminated $(J, 2J)$-regular LDPC convolutional code families converge to a value close to capacity and that this value improves as we increase $J$. This indicates that, for large $L$, both the distance growth rates and the thresholds improve with increasing complexity. We would expect this trend to continue as we further increase the variable node degree $J$, although the improvement will diminish with increasing $J$.

Now consider choosing $L$ such that the ensemble design rate is $R = 0.49$. In this region, the threshold values of the $(J, 2J)$-regular ensembles improve with $J$. The thresholds of the terminated $(3, 6)$-regular and $(4, 8)$-regular ensembles are displayed in Fig. 1, along with the simulated performance of randomly chosen codes from the associated ensembles with permutation matrix size $N = 6000$. A standard LDPC block decoder employing the BP decoding algorithm is used. We observe that the waterfall performance is relatively close to the threshold and we expect this gap to decrease for larger permutation matrix sizes $N$. By choosing $L$ larger, the rate increases (approaching $1/2$) and the thresholds move to the left. The corresponding waterfall performance of codes chosen from these ensembles will also move to the left. As a final observation, we note that for all achieveable rates the TARJA ensembles have better thresholds than the terminated $(3,6)$-regular ensembles. This is expected, since the ARJA ensemble has been optimised to have a good iterative decoding threshold. However, we also observe that, for large $L$, the $(4,8)$- and $(5,10)$-regular terminated ensembles have comparable thresholds to the TARJA ensemble, demonstrating the benefit that derives from terminating the $(J,2J)$-regular convolutional structure.

Figure 3 plots the typical minimum distance growth rates against the threshold gap to capacity (the difference between the calculated AWGN channel threshold ($E_b/N_0$) of an ensemble and capacity for the ensemble design rate) for the terminated $(J,2J)$-regular convolutional ensembles with termination factors $L = m_s + 1, \ldots, 16, 20, 50, 100$, the TARJA ensembles with termination factors $L = 2, \ldots, 10$, the ARJA block code ensemble, and the corresponding $(J,2J)$-regular block code ensembles.

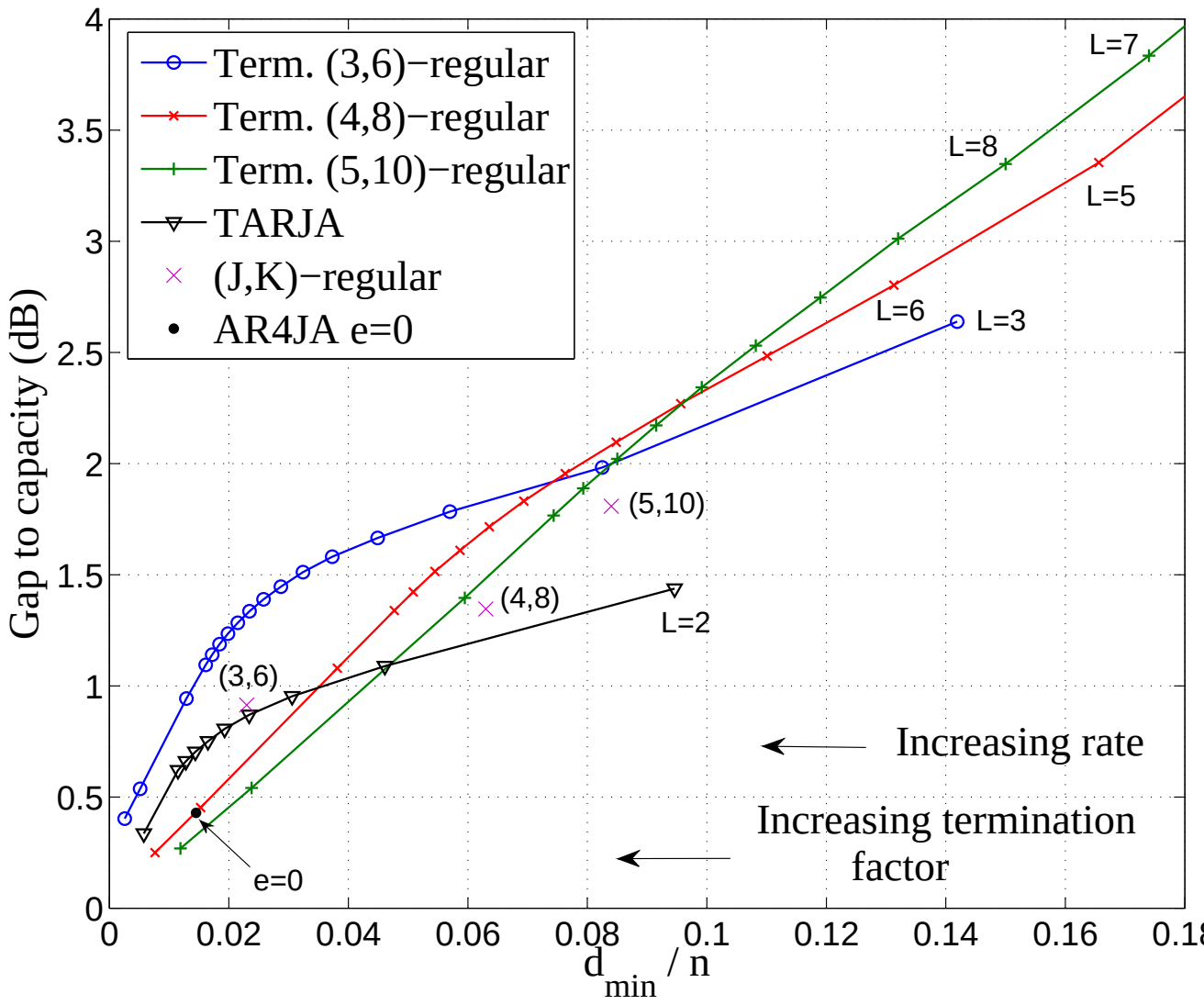

Fig. 3: Typical minimum distance growth rate vs. threshold gap to capacity.

We observe that, in particular, intermediate values of $L$ provide thresholds with a small gap to capacity while maintaining a small typical minimum distance growth rate with only a slight loss in code rate. We also note that, for a fixed gap to capacity close to zero, the largest minimum distance growth rate is obtained by choosing the terminated $(J,2J)$ ensemble with the largest $J$, and that the TARJA ensemble falls in between the terminated $(3,6)$- and $(4,8)$-regular ensembles. (In this region, with the gap to capacity close to zero, the rates are approximately equal and close to $1/2$.) For larger fixed gaps to capacity, we see that this ordering changes, and the reverse ordering holds for large gaps to capacity.

## IV. Trapping set analysis of terminated LDPC convolutional code ensembles

In [5], MacKay and Postol discovered a "weakness" in the structure of the Margulis construction of a $(3,6)$-regular Gallager code. Described as *near-codewords*, these small graphical sub-structures existing in the Tanner graph of LDPC codes cause the iterative decoding algorithm to get trapped in error patterns. These weaknesses were shown to contribute significantly to the performance of the code in the error floor region of the BER curve. Richardson developed this concept in [6], and defined these structures as *trapping sets*.

*Definition 1:* An $(a,b)$ general trapping set $\tau_{a,b}$ of a bipartite graph is a set of variable nodes of size $a$ which induce a subgraph with exactly $b$ odd-degree check nodes (and an arbitrary number of even-degree check nodes).

In order to calculate ensemble average general trapping set enumerators for terminated LDPC convolutional codes, we use the combinatorial arguments previously presented in [9]. The technique involves considering a two-part ensemble average weight enumerator for a modified protograph with the property that any $(a,b)$ trapping set in the original protograph is a codeword in the modified protograph.

### A. Trapping set growth rates

The two-part normalized logarithmic *asymptotic trapping set spectral shape function* of a code ensemble can be written as $r(\alpha,\beta) = \limsup_{n\to\infty} r_n(\alpha,\beta)$, where $r_n(\alpha,\beta) = \frac{\ln(A_{a,b})}{n}$, $\alpha = a/n$, $\beta = b/n$, $a$ and $b$ are Hamming weights, $n$ is the block length, and $A_{a,b}$ is the two-part ensemble average weight distribution. Suppose now we are interested in the ratio of $b$ to $a$ for a general $(a,b)$ trapping set enumerator. Let $\Delta = b/a = \beta/\alpha$, $\Delta \in [0,\infty)$. As proposed in [9], we may classify the trapping sets as $\tau_\Delta = \{\tau_{a,b} | b = \Delta a\}$. For each $\Delta$, we define $d_{ts}(\Delta)$ to be the $\Delta$*-trapping set number*, which is the size of the smallest, non-empty trapping set in $\tau_\Delta$. Now consider fixing $\Delta$ and plotting the normalized weight $\alpha$ against the two-part asymptotic spectral shape function $r(\alpha,\beta) = r(\alpha,\Delta\alpha)$. Suppose $\alpha > 0$ and the first zero-crossing of $r(\alpha,\beta)$ occurs at $\alpha = \delta_{ts}(\Delta)$. If $r(\alpha,\beta)$ is negative in the range $0 < \alpha < \delta_{ts}(\Delta)$, then the first zero-crossing $\delta_{ts}(\Delta)$ is called the $\Delta$*-trapping set growth rate* of the code ensemble. If $\delta_{ts}(\Delta)$ exists we can say with high probability that a randomly chosen code from the ensemble has a $\Delta$-trapping set number that is at least as large as $n\delta_{ts}(\Delta)$, i.e., the $\Delta$-trapping set number increases linearly with block length $n$ [9]. This implies that, for sufficiently large $n$, a typical member of the ensemble has no small trapping sets.

### B. Trapping set analysis of terminated $(3,6)$-regular LDPC convolutional codes

As an example, we will consider the terminated $(3,6)$-regular LDPC convolutional code ensemble described in Section II-A. For each termination factor $L$, we analyse the two-part asymptotic spectral shape function described in Section IV-A with $\Delta \geq 0$ to see if a positive trapping set growth rate $\delta_{ts}^{(L)}(\Delta)$ exists. Note that setting $\Delta = \beta/\alpha = b/a = 0$ corresponds to the minimum distance growth rate problem, where $\alpha$ and $\beta$ are the weights $a$ and $b$ normalised by the block length $n$. Thus, $\delta_{ts}^{(L)}(0) = \delta_{min}^{(L)}$, where $\delta_{min}^{(L)}$ is the minimum distance growth rate of the terminated ensemble as reported in [4]. As $\Delta$ ranges from 0 to $\infty$, the points $(\delta_{ts}^{(L)}(\Delta), \Delta\delta_{ts}^{(L)}(\Delta))$ trace out the so-called *zero-contour curve* for a protograph-based code ensemble [9]. The zero-contour curves for terminated $(3, 6)$-regular LDPC convolutional code ensembles are shown in Fig. 4 for $L = 3, \ldots, 12$.

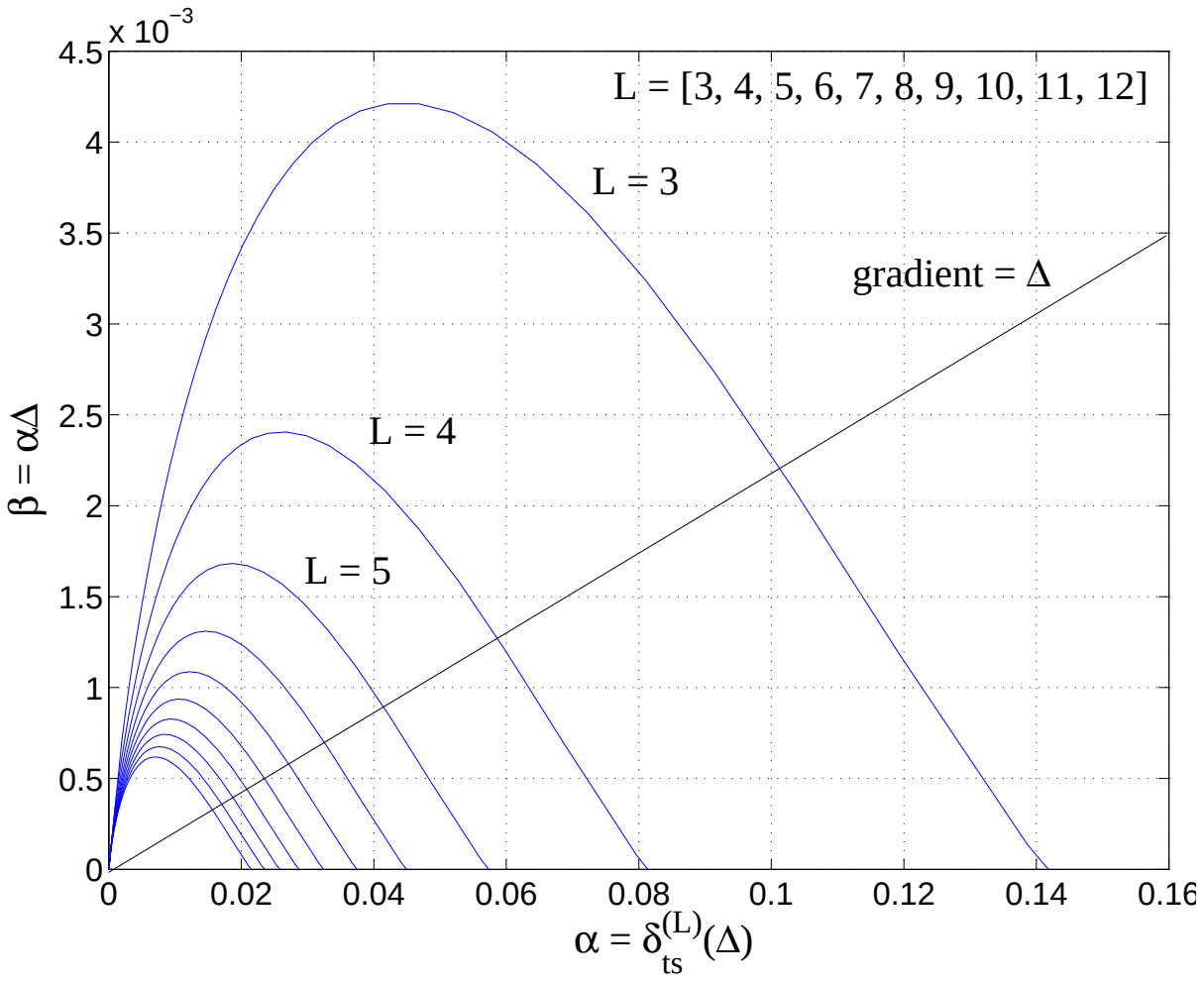

Fig. 4: Zero contour curves for terminated $(3, 6)$-regular LDPC convolutional code ensembles.

For all $\Delta \geq 0$, we observe $\delta_{ts}^{(L)}(\Delta) > 0$, indicating that, for each class of $(a, b)$ general trapping set, the size of the smallest non-empty trapping set typical of most members of the ensemble is growing linearly with block length. Code ensembles with large $\Delta$-trapping set numbers $d_{ts}^{(L)}(\Delta)$ are the most interesting, since small trapping sets dominate iterative decoding performance in the error floor [6]. Thus we want the $\Delta$-trapping set growth rate $\delta_{ts}^{(L)}(\Delta)$ to exist and to be as large as possible for each value of $\Delta$. We observe in Fig. 4 that $\delta_{ts}^{(L_1)}(\Delta) \leq \delta_{ts}^{(L_2)}(\Delta)$ for any $L_1 > L_2$. This is analogous to the decrease in the minimum distance growth rate with increasing $L$ (and rate $R$) observed in [4]. These results suggest that for larger values of $L$, where it becomes problematic to calculate the trapping set growth rates numerically, we will observe positive zero-contour curves with $\delta_{ts}^{(L)}(0) = \delta_{min}^{(L)} > 0$, the minimum distance growth rate of the terminated ensemble. This promises, for sufficiently large block length $n$, good error-floor performance for terminated $(3, 6)$-regular LDPC convolutional code ensembles in addition to the capacity approaching thresholds discussed earlier.

## V. Conclusions

In this paper we saw that the capacity approaching thresholds of terminated LDPC convolutional codes, recently established for the BEC, also extend to the AWGN channel. In addition, the terminated ensembles display linear minimum distance growth for any finite termination factor $L$. An asymptotic trapping set analysis was performed on a family of terminated $(3, 6)$-regular LDPC convolutional code ensembles and it was shown that they possess the property that the smallest non-empty trapping set grows linearly with the block length. These properties indicate that codes chosen from these ensembles should have excellent performance in both the waterfall and the error-floor region of the BER curve.

## Acknowledgements

The authors would like to thank Maria Mellado Prenda for providing the simulation curves. This work was partially supported by NSF Grant CCF08-30650 and NASA Grant NNX09-AI66G.